\documentclass[twocolumn,prd,longbibliography,
 amsmath,amssymb,superscriptaddress]{revtex4-2}
\usepackage{graphicx}
\usepackage{dcolumn}
\usepackage{bm}
\usepackage[colorlinks=true, citecolor=blue, linkcolor=blue, urlcolor=blue]{hyperref}
\usepackage{xcolor}
\usepackage{amsmath, amssymb, amsfonts}
\usepackage{mathtools}
\usepackage{subfigure}
\usepackage{mathrsfs}
\usepackage{sidecap}
\usepackage{verbatim}

\definecolor{purple1}{rgb}{128,0,128}

\usepackage{bigints}
\newcommand{\bea}{\begin{eqnarray}}
\newcommand{\ea}{\end{eqnarray}}

\definecolor{darkpastelgreen}{rgb}{0.01, 0.75, 0.24}

\def\d{\mathrm{d}}

\begin{document}

\title{
Energy conservation and quantum backreaction in Bose-Einstein condensates} 
\author{Caio C. \surname{Holanda Ribeiro}}
\email{caiocesarribeiro@alumni.usp.br}
\affiliation{International Center of Physics, Institute of Physics, University of Brasilia, 70297-400 Brasilia, Federal District, Brazil} 

\date\today

\begin{abstract}
Bose-Einstein condensates are suitable systems for studying fundamental aspects of quantum backreaction. Here the backreaction problem in 1D condensates is considered from the perspective of energy and momentum conservation. By assuming the validity of Bogoliubov theory, the backreaction equations are used to identify the contributions to the system energy and momentum coming from quantum fluctuations and condensate corrections. The backreaction 
is solved for a condensate trapped in a ring configuration and such that particle interactions are continuously switched on. It is shown that the energy in the condensate cannot be addressed without taking into account how the system entered the interacting regime, and even for homogeneous condensates the power transferred to the condensate by quantum fluctuations showcases an intricate non-monotonic pattern.

\end{abstract}

\maketitle

\section{Introduction}

The problem of determining how backreaction occurs in semiclassical gravity is one of the most intricate and rewarding ones in quantum field theory. In simple terms, the problem consists in determining how quantum fluctuations of test fields on top of a classical curved background influence the spacetime. Examples of quantum backreaction effects include the spontaneous Hawking radiation by black holes, which is believed to diminish the black hole mass \cite{HAWKING1974}, and certain chronological protection mechanisms in cosmic string spacetimes \cite{vitorio1,vitorio2}. Other interesting example is known as the gravity-induced vacuum dominance \cite{daniel1}, where the quantum fluctuations become sufficiently strong and force the spacetime into a different configuration via decoherence \cite{daniel2,daniel3}. Yet, despite the plethora of interesting examples, some backreaction problems of fundamental importance remain not fully solved \cite{daniel3}. We cite, for instance, the problem of determining how quantum fluctuations affect gravitational collapse in general. Only partial results exist \cite{Ziprick,Shafiee2023}, and they suggest that quantum fluctuations can play a prominent role on black hole formation. 

Another difficulty in studying backreaction in semiclassical gravity is related to the available experimental data, which, for some weak effects like the Hawking radiation, might be virtually impossible to obtain. One possible route for studying such small effects is offered by analogue gravity  in Bose-Einstein condensates (BEC). This program, that gained momentum with \cite{Unruh1981}, is well-developed \cite{Visser2011}, and counts with examples ranging from black hole to cosmological analogues \cite{Garay,Fischer2004,Jain,Steinhauer2016,Jacobson2018}. Also, the analogue gravity community has achieved an impressive milestone: The measurement of an analogue Hawking radiation \cite{Jeff2019,Jeff2021}. BECs are also systems that can be used to study some quantum backreaction effects \cite{Balbinot2005,Liberati2020,Iacopo2023}. Within the analogue gravity program, we cite the possibility of simulating the evolution of quantum fluctuations on top of analogue black hole spacetimes \cite{Jacobson2017,Cesar}, one of the necessary ingredients to find the corresponding corrections to the background analogue spacetime.

Different methods can be implemented in order to study quantum backreaction in BECs. In an ideal scenario, one seeks a solution for the full non-linear dynamics of a gas of bosons trapped in some region and subjected to initial conditions. This goal, however, cannot be fulfilled in general due to the elevated complexity of the field equations. Yet, in some cases, when it is possible to identify small quantum fluctuations, approximate solutions can be found by means of the Bogoliubov expansion or other equivalent methods \cite{castin,gardiner}. This idea was explored in \cite{Fischer2005}, where the authors developed an approach to the quantum backreaction in BECs based on the particle number conservation. This method was recently applied for a concrete condensate configuration \cite{RibeiroPRA} and a solution to the quantum backreaction was found.  

In this work we go a step further in the analysis initiated in \cite{RibeiroPRA} by discussing another crucial feature of the backreaction analysis, which is the energy conservation. By starting with a non-relativistic gas of bosons allowed to move in one spatial dimension, we implement an expansion of the field operator in powers of the number of particles in the system, thus identifying a notion of background condensate, quantum fluctuations, and condensate corrections. We then adopt the canonical stress tensor, $T^{\mu}_{\ \nu}$, as a measure of the system energy and momentum densities, and identify the different contributions to $T^{\mu}_{\ \nu}$ coming from the quantum fluctuations and the condensate corrections, as well as the associated conservation laws. It is shown that the backreaction scheme of \cite{Fischer2005,RibeiroPRA} can be restated in terms of one of the conservation laws linked to $T^{\mu}_{\ \nu}$, namely, the conservation of momentum. The latter is capable of revealing the role played by the various contributions of the Euler equation of \cite{RibeiroPRA}, pinpointing the sources of stresses in the system and forces due to external agents. 

The backreaction equations are solved for a homogeneous condensate trapped in a ring configuration which undergoes a quantum quench from an initially non-interacting regime to an interacting one. This is important in order to avoid the vacuum state indeterminacy coming from the phenomenon of phase diffusion \cite{lew}. In order to keep the model experimentally appealing, we employ a continuous transition between the different regimes which contains the sudden transition as a particular case and such that quantization can be performed exactly. The solution to the backreaction problem is then used to fully characterize the energy of the system.

This work is organized as follows. In Sec.~\ref{section2} the general description of the non-relativistic gas of bosons is reviewed, and the backreaction equations are obtained. In Sec.~\ref{section3} the energy-momentum tensor is presented and its various contributions are identified, as well as the corresponding conservation laws. Section \ref{section4} discusses in details the construction of a quantum field theory for the quantum fluctuations in the system. In particular, the connection between the system vacuum state and different instantaneous quasiparticle vacua is presented. In Sec.~\ref{section5} the condensate depletion and the system energy are thoroughly studied, and we finish with some final remarks in Sec.~\ref{section6}.

\section{BEC Lagrangian and field equations}
\label{section2}

We consider a non-relativistic gas of bosons allowed to move in only one dimension, hereafter parameterized by $x$. We let $\phi=\phi(t,x)$ be the complex field describing the gas according to the Lagrangian (in units of $\hbar=1$)
\begin{equation}
    \mathcal{L}=i\phi^*\partial_t\phi-\frac{1}{2m}|\partial_x\phi|^2-\left(V+\frac{g}{2}|\phi|^2\right)|\phi|^2,\label{fulllagrangian}
\end{equation}
where in general $V=V(t,x), g=g(t,x)$, $m,g>0$, with $m$ being a constant. Note that the above Lagrangian is invariant under the global $U(1)$ transformation $\phi\rightarrow \exp(i\alpha)\phi$, for any constant real parameter $\alpha$. This in turn implies by means of Noether's Theorem the local conservation law \cite{Maggiore}
\begin{equation}
    \partial_t\rho+\partial_xJ=0,\label{numberconservation}
\end{equation}
where
\begin{align}
\rho&=|\phi|^2,\\
J&=\frac{1}{m}\mbox{Im}(\phi^*\partial_x\phi).
\end{align}
Therefore $\rho$ can be interpreted as the particle density, with
\begin{equation}
    N=\int\d x\rho,
\end{equation}
being the total number of gas particles in the system. Finally, by treating $\phi$ and $\phi^*$ as independent variables, the Euler-Lagrange equation for $\phi^*$ reads
\begin{equation}
    i\partial_t\phi=-\frac{\partial^2_x\phi}{2m}+(V+g|\phi|^2)\phi.\label{fieldeq}
\end{equation}
%
%
%

As in any quantum field theory, given the system parameters $m,g,V$, the quantum features of the system can be studied by looking for an operator-valued distribution, $\phi$, solution of Eq.~\eqref{fieldeq}, such that
\begin{equation}
    \left[\phi(t,x),\pi(t,x')\right]=i\delta(x-x')\label{ccrtot}
\end{equation}
holds true, where %
\begin{equation}
    \pi=i\phi^\dagger
\end{equation}
is the momentum canonically conjugate to $\phi.$

There exists a plethora of possible routes to quantize the above theory, and here we adopt the Bogoliubov expansion \cite{Fischer2005}. Specifically, we are interested in the regime $N\gg 1$, while $g N\sim 1$. When this occurs it is possible to separate the field $\phi$ into a ``classical'' part plus ``small'' quantum fluctuations by expanding $\phi$ in powers of $\sqrt{N}$ as
\begin{equation}
    \phi=\phi_0+\chi+\zeta+\ldots,\label{expansion}
\end{equation}
where formally $\phi_0$ is proportional to $\sqrt{N}$, $\chi$ is proportional to $N^{0}$, and $\zeta$, to $N^{-1/2}$. Thus it follows from this ansatz and the Euler-Lagrange equation that $\phi_0$ is solution of the Gross-Pitaevskii equation \cite{RibeiroPRA},
\begin{align}
i\partial_t&\phi_{0}=\left(-\frac{\partial_x^2}{2m}+V+g\rho_0\right)\phi_{0},\label{GP}
\end{align}
with $\rho_0=|\phi_0|^2$, $\chi$ is ruled by the  
the Bogoliubov-de Gennes (BdG) equation \cite{RibeiroPRA}
\begin{align}
i\partial_t\chi=\left(-\frac{\partial_x^2}{2m}+V+2g\rho_0\right)\chi+g\phi_{\rm 0}^2\chi^{*},\label{BdG}
\end{align}
and finally a  BdG-like equation with $\chi$-dependent  
source terms for $\zeta$ \cite{RibeiroPRA}, 
\begin{align}
i\partial_t\zeta=&\left(-\frac{\partial_x^2}{2m}+V+2g\rho_0\right)\zeta+g\phi_{\rm 0}^2\zeta^{*}\nonumber\\
&+2g|\chi|^2\phi_0+g\chi^2\phi_0^*,\label{GPcorr}
\end{align} 
appears at order $N^{-1/2}$. We stress that it is crucial to keep the field $\zeta$ in the expansion in order to ensure that quantum fluctuations are correctly accounted for in the conservation laws, as we discuss in the next section.

Within the Bogoliubov expansion, which holds true as long as the hierarchy in separation \eqref{expansion} is preserved, the quantum properties of theory are assumed to be encapsulated in the field $\chi$ subjected, by means of Eq.~\eqref{ccrtot}, to the canonical commutation relation
\begin{equation}
    \left[\chi(t,x),\chi^\dagger(t,x')\right]=\delta(x-x').
\end{equation}
Physically, the above theory describes condensates composed of a sizable background described by $\phi_0$, and on top of which small quantum fluctuations, described by $\chi$, exist. The field $\zeta$ then models eventual corrections to the background condensate. This is the idea behind the quantum backreaction discussed in \cite{Fischer2005,RibeiroPRA}.

\section{Energy-momentum tensor and conservation laws}
\label{section3}

We now turn our attention to the conservation laws associated to the system particle number, energy, and momentum within the quantum backreaction scheme presented in the previous section. The particle number conservation was briefly discussed in Eq.~\eqref{numberconservation} in order to identify number of particles, $N$, of the gas. For the system energy and momentum we adopt the canonical energy-momentum (stress) tensor as their measure. At the classical level, by letting $\phi=\phi_1$, $\phi^{*}=\phi_2$, and $x^{\mu}=(t,x)$, the  canonical energy-momentum tensor, $T^{\mu}_{\ \nu}$, for a given Lagrangian $\mathcal{L}$ is defined by \cite{Maggiore}
\begin{equation}
    T^{\mu}_{\ \nu}=\frac{\partial\mathcal{L}}{\partial(\partial_\mu\phi_i)}\partial_\nu\phi_i-\delta^{\mu}_{\ \nu}\mathcal{L}.
\end{equation}
Therefore, it follows from Eq.~\eqref{fulllagrangian}  and the Euler-Lagrange equation that
\begin{equation}
    \partial_\mu T^{\mu}_{\ \nu}=|\phi|^2\left(\frac{|\phi|^2}{2}\partial_\nu g+\partial_\nu V\right),\label{cce}
\end{equation}
where 
\begin{align}
    T^{0}_{\ 0}&=\frac{1}{2m}|\partial_x\phi|^2+\left(V+\frac{g}{2}|\phi|^2\right)|\phi|^2,\label{t1}\\
    T^{1}_{\ 0}&=-\frac{1}{2m}(\partial_x\phi^*)\partial_t\phi-\frac{1}{2m}(\partial_t\phi^*)\partial_x\phi,\label{t2}\\
    T^{0}_{\ 1}&=i\phi^*\partial_x\phi,\label{t3}\\
    T^{1}_{\ 1}&=-i\phi^*\partial_t\phi-\frac{1}{2m}|\partial_x\phi|^2+\left(V+\frac{g}{2}|\phi|^2\right)|\phi|^2.\label{t4}
\end{align}
We thus immediately recognize $T^{0}_{\ 0}$ to be the system Hamiltonian density obtained from the Lagrangian \eqref{fulllagrangian}, which in turn implies that $T^{1}_{\ 0}$ is the system energy flux, whereas $-\mbox{Re}(T^{0}_{\ 1})=mJ$ is the system momentum density, and $\mbox{Re}(T^{1}_{\ 1})$ is the corresponding system stress \cite{wald2022}.

We note also that if $\partial_\nu g=0$ and $\partial_\nu V=0$, then Eq.~\eqref{cce} becomes a continuity equation: $\partial_tT^{0}_{\ \nu}+\partial_xT^{1}_{\ \nu}=0$ and the corresponding ``charge'' $\int\d x T^{0}_{\ \nu}$ is conserved.

In order to establish the conservation laws at the quantum level it is necessary to adopt some ordering prescription when building operators from the classical observable quantities, e.g, $\rho=|\phi|^2$. Indeed, as $\chi$ and $\chi^\dagger$ are noncommuting variables, in general $\langle\chi^\dagger\chi\rangle\neq\langle\chi\chi^\dagger\rangle$. This affects, for instance, the evolution of $\zeta$, which depends on products of $\chi$ and $\chi^\dagger$. In this work we adopt the normal ordering prescription, which corresponds to define quantum operators by putting all $\chi^\dagger$ to the left of all $\chi$ in all monomials. Also, we assume that $\chi$ is prepared in a given vacuum state, for which $\langle \chi\rangle=0$. 

With the normal ordering prescription, the c-number $\zeta$ is taken to satisfy the vacuum expectation of Eq.~\eqref{GPcorr}
\begin{align}
i\partial_t\zeta=&\left(-\frac{\partial_x^2}{2m}+V+2g\rho_0\right)\zeta+g\phi_{\rm 0}^2\zeta^{*}\nonumber\\
&+2g\langle\chi^\dagger\chi\rangle\phi_0+g\langle\chi^2\rangle\phi_0^*,\label{GPcorr2}
\end{align} 
thus fixing how the condensate is corrected by the quantum fluctuations.

The first conservation law we discuss is the one given by Eq.~\eqref{numberconservation}. Upon quantization, we redefine the gas density to be $\rho=\langle\phi^\dagger\phi\rangle$ and the particle flux as $J=(1/m)\mbox{Im}(\langle\phi^\dagger\partial_x\phi\rangle)$. Thus the expansion \eqref{expansion} implies \cite{RibeiroPRA}
\begin{align}
    \rho&=\rho_0+\rho_\chi+\rho_\zeta+\mathcal{O}(N^{-1}),\\
    J&=J_0+J_\chi+J_\zeta+\mathcal{O}(N^{-1}),
\end{align}
where $J_0=(1/m)\mbox{Im}(\phi^*_0\partial_x\phi_0)$ is the background particle flux,
\begin{align}
    \rho_\chi&=\langle\chi^\dagger\chi\rangle,\\
    J_\chi&=\frac{1}{m}\mbox{Im}(\langle\chi^\dagger\partial_x\chi\rangle),
\end{align}
are the density of depleted particles and the phonon flux, and 
\begin{align}
\rho_{\zeta}&=2\mbox{Re}\ [\phi_0^*\zeta],\label{rhozeta1}\\
J_{\zeta}&=\frac{1}{m}\mbox{Im}\ [\phi_0^*\partial_x\zeta+(\partial_x\phi_0)\zeta^*],\label{jzeta}
\end{align}
are the corrections to the condensate density and particle flux, respectively. The importance of keeping the field $\zeta$ in the expansion resides in the fact that $\partial_t\rho_0+\partial_xJ_0=0$ always, whereas \cite{RibeiroPRA}
\begin{subequations}
\label{continuity}
\begin{align}
&\partial_t\rho_{\chi}+\partial_{x}J_{\chi}=-ig(\phi_{0}^2\langle\hat{\chi}^{\dagger2}\rangle-\phi_{0}^{*2}\langle\hat{\chi}^2\rangle),\label{cont2}\\
&\partial_t\rho_{\zeta}+\partial_{x}J_{\zeta}=ig(\phi_{0}^2\langle\hat{\chi}^{\dagger2}\rangle-\phi_{0}^{*2}\langle\hat{\chi}^2\rangle),\label{cont1}
\end{align}
\end{subequations}
which is the mathematical expression for the breakdown of particle number conservation by the Bogoliubov expansion. Although the number of depleted particles alone is not conserved, $\partial_t(\rho_\chi+\rho_\zeta)+\partial_x(J_\chi+J_\zeta)=0$. A detailed account of these various contributions was recently considered in \cite{RibeiroPRA}.

Following the normal ordering procedure, we define the system energy density as $\mathcal{E}=\langle:T^{0}_{\ 0}:\rangle$, the system energy flux by $S=\langle:T^{1}_{\ 0}:\rangle$, the system momentum density $\mathcal{P}=-\mbox{Re}(\langle:T^{0}_{\ 1}:\rangle)$, and $\Theta=\mbox{Re}(\langle:T^{1}_{\ 1}:\rangle)$ is the system stress. Thus, the expansion \eqref{expansion} implies
\begin{align}
    \mathcal{E}&=\mathcal{E}_0+\mathcal{E}_\chi+\mathcal{E}_\zeta+\mathcal{O}(N^{-1}),
\end{align}
and similarly for $S, \mathcal{P}$, and $\Theta$. The dominant (order $N$) contributions, $\mathcal{E}_0$, $S_0$, $\mathcal{P}_0$, $\Theta_0$ are obtained from Eqs.~\eqref{t1}, \eqref{t2}, \eqref{t3}, and \eqref{t4} by making the substitution $\phi\rightarrow\phi_0$, whereas for the remaining terms, we find
\begin{align}
    \mathcal{E}_\chi&=
    \frac{1}{2m}\langle(\partial_x\chi^\dagger)\partial_x\chi\rangle+(V+g\rho_0)\rho_\chi+\frac{g}{2}G^{(2)},\label{energychi}\\
    \mathcal{E}_\zeta&=\frac{1}{m}\mbox{Re}\left[(\partial_x\phi^*_0)\partial_x\zeta\right]+(V+g\rho_0)\rho_\zeta,\label{energyzeta}
\end{align}
which are identified as the energy density of the depleted cloud and the correction to the background condensate energy density, respectively. Here, $G^{(2)}=\langle:(\rho-\langle\rho\rangle)^2:\rangle$ is the local density variance. For the energy flux, it follows from Eq.~\eqref{t2} that
\begin{align}
    S_\chi&=\frac{1}{m}\mbox{Re}[\langle(\partial_x\chi^\dagger)\partial_t\chi\rangle],\\
    S_\zeta&=\frac{1}{m}\mbox{Re}[(\partial_x\zeta^*)\partial_t\phi_0+(\partial_x\phi_0^*)\partial_t\zeta].
\end{align}
Thus, the general equation \eqref{cce} implies, for $\nu=0$, that
\begin{align}
    \partial_t(\mathcal{E}_\chi+\mathcal{E}_\zeta)+&\partial_x(S_\chi+S_\zeta)=(\partial_t V)(\rho_\chi+\rho_\zeta)\nonumber\\
    &+(\partial_tg)\left[\rho_0(\rho_\chi+\rho_\zeta)+G^{(2)}/2\right].\label{energycon}
\end{align}
The above equation, which rules the energy exchange between condensate and depleted cloud, holds at order $N^{0}$, and it shows, in particular, that for $\partial_tV=\rho_0\partial_tg=0$, the energy $\int\d x(\mathcal{E}_\chi+\mathcal{E}_\zeta)$ is conserved for confined condensates.

The last conservation law is the one linked to the system momentum, $\mathcal{P}$. We find that $\mathcal{P}=mJ=m(J_0+J_\chi+J_\zeta)$, and
\begin{align}
    \Theta_\chi&=\frac{\partial^2_x\rho_\chi}{4m}-\frac{1}{m}\langle(\partial_x\chi^\dagger)\partial_x\chi\rangle,\\
    \Theta_\zeta&=\mbox{Im}[\zeta^*\partial_t\phi_0+\phi_0^*\partial_t\zeta]-\frac{1}{m}\mbox{Re}[(\partial_x\phi_0^*)\partial_x\zeta]\nonumber\\
    &\ \ \ \ +(V+g\rho_0)\rho_\zeta.
\end{align}
Therefore, Eq.~\eqref{cce} for $\nu=1$ implies
\begin{align}
    \partial_t(\mathcal{P}_\chi+\mathcal{P}_\zeta)-&\partial_x(\Theta_\chi+\Theta_\zeta)=-(\partial_x V)(\rho_\chi+\rho_\zeta)\nonumber\\
    &-(\partial_xg)\left[\rho_0(\rho_\chi+\rho_\zeta)+G^{(2)}/2\right].\label{momentumcon}
\end{align}
The right hand side of the above equation is the force density applied on the system by external laboratory agents. Naturally, for $\partial_xV=\rho_0\partial_xg=0$, the total momentum $\int\d x(\mathcal{P}_\chi+\mathcal{P}_\zeta)$ is conserved in confined condensates.

We conclude this section with some remarks regarding the connection of the conservation laws \eqref{energycon} and \eqref{momentumcon} with previous works. Equation \eqref{momentumcon} is essentially the Euler-like equation upon which the backreaction analysis of \cite{RibeiroPRA} is based. The novelty of Eq.~\eqref{momentumcon} is that it identifies the sources of the various contributions to the system momentum and stress. Furthermore, although in principle it is possible to determine the evolution of $J_{\zeta}$ and $\rho_\zeta$ once the quantum fluctuations on the system are known, it is not clear, from the results in \cite{RibeiroPRA}, how to determine the system energy from these data, in a similar fashion to what occurs to the determination of the electromagnetic stress tensor from Maxwell's equations \cite{wald2022}. We recall that at a fundamental level the true stress tensor of a given field theory is the one that sources gravity through Einstein's equations. Nevertheless, although the canonical stress tensor might not coincide with the true stress tensor, for our non-relativistic case the system energy and momentum are indeed correct, and  Eqs.~\eqref{energychi} and \eqref{energyzeta} give a clear separation between the energy density of the depleted cloud and of the corresponding correction to the condensate, whereas Eq.~\eqref{energycon} expresses how the latter evolve in the condensate.

\section{Homogeneous condensates in a ring}
\label{section4}

In this section we present the quantization of $\chi$ for a family of condensates of great experimental appeal: Homogeneous condensates, at rest with respect to the laboratory frame, trapped in ring configurations. It is important to notice that some works treat condensates in the thermodynamic limit, which are idealized infinitely extended systems. For instance, in \cite{Fischer2005} it was shown how the condensate density is corrected for a 3D gas in the thermodynamic limit. However, in actual experiments it is not completely clear how to decide whether or not the thermodynamic limit is a good approximation, and for finite condensate configurations phase diffusion might play a prominent role. 

We consider a trapping potential for which $\phi$ appearing in Eq.~\eqref{fulllagrangian} is defined for $-\ell/2<x\leq\ell/2$, and the following boundary conditions
\begin{align}
\left.\phi\right|_{x=-\ell/2}&=\left.\phi\right|_{x=\ell/2},\\
\left.\partial_x\phi\right|_{x=-\ell/2}&=\left.\partial_x\phi\right|_{x=\ell/2},
\end{align}   
are naturally imposed by the field equation. Accordingly, because of their independence at each order in the expansion in powers of $N^{1/2}$, the fields $\phi_0$, $\chi$, and $\zeta$ are also subjected to the same conditions.

Another prominent aspect in any quantum backreaction analysis in condensates is the difficulty in choosing a quantum vacuum state that can describe a given experimental realization. This difficulty emerges from the fact that the field $\chi$ can be dynamically unstable, as occurs for instance in the phenomenon of gravity-induced instability \cite{daniel1}, or it might describe a condensate undergoing phase diffusion \cite{lew}. In both cases, even when the background condensate is stationary, the quantum fluctuations on top of it can grow, leading to the spontaneous breakdown of the time-translation symmetry and the consequent difficulty in knowing the condensate state at future instants of time. In such scenarios, it is necessary to provide extra assumptions in order to select a meaningful vacuum state.     

A homogeneous condensate is thus obtained by assuming a constant potential $V$ and a background order parameter $\phi_0=\sqrt{\rho_0}\exp(-i\mu t)$, for constant $\rho_0$ and $\mu$, such that the GP equation \eqref{GP} reduces to an equation for the chemical potential $\mu$:
\begin{equation}
\mu=V+g\rho_0.\label{chemical}
\end{equation}   
An interesting strategy to avoid the indeterminacy of the system vacuum is to assume that the condensate under consideration starts from a non-interacting regime ($g=0$), say, for $t<0$, and then reaches an interacting regime after some switching time $\tau_s>0$, for which $g=g_0>0$ becomes constant. Thus, $\mu=V\equiv\mu_0$ for $t<0$ and $\mu=V+g_0\rho_0\equiv V+\Delta\mu$ for $t>\tau_s$, where $\Delta\mu=g_0\rho_0$. During the switching time, we assume that the interactions are linearly turned on, according to
\begin{equation}
    g(t)=g_0\left\{
    \begin{array}{cc}
       0,  & t<0, \\
       t/\tau_s,  & 0\leq t\leq \tau_s,\\
       1, & \tau_s<t.
    \end{array}\right.\label{eqforg}
\end{equation}
Henceforth we call the periods $t<0$, $0\leq t\leq\tau_s$, $\tau_s<t$ the non-interacting regime, the switching regime, and the interacting regime, respectively. This profile has the advantage of allowing for the construction of exact solutions.

We call attention to the spatial translation invariance of this condensate realization, which implies that there are no particle fluxes during the condensate evolution, i.e., $J_0=J_\chi=J_\zeta=0$, and the continuity equation reduces to $\partial_t(\rho_\chi+\rho_\zeta)=0$. Accordingly, $\rho_\zeta=-\rho_\chi$, and the condensate density is determined by the quantum depletion alone. We stress that this property does not occur for more complex condensate configurations. We cite \cite{RibeiroPRA} for the study of a case where the particle fluxes are relevant.

\subsection{Field quantization}

In order to find a suitable quantum field expansion for $\chi$, we start by defining a new variable $\psi(t,x)=\exp(i\mu t)\chi(t,x)$, such that Eq.~\eqref{BdG} reduces to
\begin{equation} 
i\partial_t\psi=-\frac{\partial_x^2\psi}{2m}+g\rho_0(\psi+\psi^*).\label{BdGred}
\end{equation}
The solutions of this equation are easily found in terms of the Nambu spinor $\Phi=(\psi, \psi^*)^{T}$, where $T$ stands for the matrix transpose. The spinor $\Phi$ satisfies
\begin{equation} 
\frac{i}{\Delta \mu}\sigma_3\partial_t\Phi=-\frac{\xi_0^2\partial_x^2\Phi}{2}+\frac{g}{g_0}(1+\sigma_1)\Phi,\label{BdGspinor}
\end{equation}
where $\xi_0=1/\sqrt{mg_0\rho_0}$ is the condensate healing length in the interacting regime and $\sigma_i$, $i=1,2,3$, are the Pauli matrices. In order to keep the equations simpler, we assume units for $x$ and $t$ such that $\xi_0=\Delta\mu=1$. At the end of the calculations units are covered by making $x\rightarrow x/\xi_0$ and $t\rightarrow t\Delta \mu$.

Only the solutions $\Phi$ of Eq.~\eqref{BdGspinor} that satisfy $\Phi=\sigma_1\Phi^*$ correspond to solutions of Eq.~\eqref{BdGred}, and for any two such $\Phi$, say $\Phi$ and $\Phi'$, \eqref{BdGspinor} implies that the scalar product
\begin{equation}
\langle\Phi,\Phi'\rangle=\int \d x\Phi^\dagger\sigma_3\Phi'
\end{equation} 
is time-independent, and can be used to normalize the field modes. If $\{\Phi_n\}_n$ is a complete set of (positive norm) field modes such that $\langle\Phi_n,\Phi_{n'}\rangle=\delta_{n,n'}$, then the general solution of Eq.~\eqref{BdGspinor} subjected to the reflection property $\Phi=\sigma_1\Phi^*$ is given by
\begin{equation}
\Phi(t,x)=\sum_n\left[a_n\Phi_n(t,x)+a_n^{*}\sigma_1\Phi_n^*(t,x)\right].\label{fieldexp}
\end{equation}

In order to find the complete set of positive norm field modes explicitly, we note first that due to the spatial translation symmetry we can look for solutions in the form 
$\Phi(t,x)=\exp(ikx)\Phi_k(t)$. Then by imposing the periodic boundary conditions, we find that $\sin(k\ell/2)=0$, from which we obtain the condition
\begin{equation}
k\equiv k_n=\frac{2n\pi}{\ell}, \ \ \ n\ \mbox{integer}.
\end{equation}

Now for each $n$, let us determine $\Phi_{k_n}(t)$. Substitution of $\Phi(t,x)=\exp(ik_nx)\Phi_{k_n}(t)$ back into Eq.~\eqref{BdGspinor} implies
\begin{equation}
    i\partial_t\sigma_3\Phi_{k_n}=\omega_n\Phi_{k_n}+\frac{g}{g_0}(1+\sigma_1)\Phi_{k_n},\label{bdgtime}
\end{equation}
where
\begin{equation}
    \omega_n=\frac{k_n^2}{2}.
\end{equation}

In the non-interacting regime ($t<0, g=0$), solutions in the form $\Phi_{k_n}(t)=\exp(-i\omega_n t)\Phi^{0}_{k_n}$ for constant $\Phi^{0}_{k_n}$ exist. In fact, Eq.~\eqref{bdgtime} fixes
\begin{equation}
\Phi^{0}_{k_n}=(1, 0)^{T},
\end{equation}
and the set of all
\begin{equation}
\Phi_n(t,x)=\frac{1}{\sqrt{\ell}}e^{-i\omega_n t+ik_n x}(1, 0)^{T},
\end{equation}
for all integer $n$ exhausts all positive norm field modes in the non-interacting regime. 

Our goal now is to determine how each $\Phi_n(t,x)$ evolves into the interacting regime ($t>0$) and this can be done in a straightforward manner by solving Eq.~\eqref{bdgtime} in terms of special functions. It is instructive to solve for the mode with $n=0$ first, as it is given in terms of elementary functions. We find that
\begin{align}
    &\Phi_0(t,x)=\frac{1}{\sqrt{\ell}}\times\nonumber\\
    &\left\{
    \begin{array}{cc}
       (1,0)^{T},  & t<0, \\
        \frac{1}{2}(1,1)^{T}+\frac{1}{2}\left(1-\frac{it^2}{\tau_s}\right)(1,-1)^{T}, & 0\leq t\leq\tau_s,\\
       \frac{1+i\tau_s-2it}{2}(1,-1)^{T}+\frac{1}{2}(1,1)^{T}, & \tau_s\leq t.
    \end{array}\right.\label{zeromode}
\end{align}
The solution for $\Phi_n$, with $n\neq0$, can be found as follows. Two linearly independent solutions of Eq.~\eqref{bdgtime} for $0<t<\tau_s$ are
\begin{align}
    &\Gamma^{(1)}_n=\frac{1}{\sqrt{l}}\left(\begin{array}{cc}
      A_i(\sigma_n)+(\frac{i}{\omega_n})(\frac{-2\omega_n}{\tau_s})^{1/3}A'_{i}(\sigma_n)  \\
      A_i(\sigma_n)-(\frac{i}{\omega_n})(\frac{-2\omega_n}{\tau_s})^{1/3}A'_{i}(\sigma_n)
    \end{array}\right),\\
    &\Gamma^{(2)}_n=\frac{1}{\sqrt{l}}\left(\begin{array}{cc}
      B_i(\sigma_n)+(\frac{i}{\omega_n})(\frac{-2\omega_n}{\tau_s})^{1/3}B'_{i}(\sigma_n)  \\
      B_i(\sigma_n)-(\frac{i}{\omega_n})(\frac{-2\omega_n}{\tau_s})^{1/3}B'_{i}(\sigma_n)
    \end{array}\right),
\end{align}
where $\sigma_n=\sigma_n(t)$ is given by
\begin{equation}
    \sigma_n(t)=-\omega_n\left(-\frac{\tau_s}{2\omega_n}\right)^{2/3}\left(\omega_n+\frac{2t}{\tau_s}\right),
\end{equation}
and $A_i(y)$, $B_i(y)$ are the two Airy functions \cite{Gradius}, solutions of Airy's equation
\begin{equation}
    \frac{\d^2A_i}{\d y^2}-yA_i=0.
\end{equation}
As for the regime $\tau_s<t$, two independent solutions for \eqref{bdgtime} are 
\begin{align}
    \Psi^{(1)}_n=\frac{e^{-i\Omega_n t}}{\sqrt{\ell(\Omega_n-\omega_n)(2-\Omega_n+\omega_n)}}\left(\begin{array}{cc}
      1  \\
      \Omega_n-\omega_n-1
    \end{array}\right),\label{psidef}
\end{align}
and $\Psi^{(2)}_n=\sigma_1\Psi^{(1)*}_n$, where
\begin{equation}
    \Omega_n=\sqrt{2\omega_n}\left(1+\frac{\omega_n}{2}\right)^{1/2}.
\end{equation}
%
Therefore, we find that
\begin{align}
    &\Phi_n(t,x)=e^{ik_nx}\left\{
    \begin{array}{cc}
       \ell^{-1/2}e^{-i\omega_n t}(1,0)^{T},  & t<0, \\
        \gamma^{(1)}_n\Gamma^{(1)}_n(t)+\gamma^{(2)}_n\Gamma^{(2)}_n(t), & 0\leq t\leq\tau_s,\\
       \eta^{(1)}_n\Psi^{(1)}_n(t)+\eta^{(2)}_n\Psi^{(2)}_n(t), & \tau_s\leq t.
    \end{array}\right.
\end{align}
The various coefficients in the above equation can be found by noticing that Eq.~\eqref{bdgtime} requires that $\Phi_n$ be a continuous function of time. Therefore, we find that
\begin{align}
    \gamma_n^{(1)}&=\frac{\Gamma^{(2)T}_{n}(0)i\sigma_2(1,0)^{T}}{\Gamma^{(2)T}_{n}(0)i\sigma_2\Gamma^{(1)T}_{n}(0)},\\
    \gamma_n^{(2)}&=\frac{\Gamma^{(1)T}_{n}(0)i\sigma_2(1,0)^{T}}{\Gamma^{(1)T}_{n}(0)i\sigma_2\Gamma^{(2)T}_{n}(0)},\\
    \eta_n^{(1)}&=\frac{\Psi^{(2)T}_{n}(\tau_s)i\sigma_2\left[\gamma^{(1)}_n\Gamma^{(1)}_n(\tau_s)+\gamma^{(2)}_n\Gamma^{(2)}_n(\tau_s)\right]}{{\Psi^{(2)T}_{n}(\tau_s)i\sigma_2\Psi^{(1)}_{n}(\tau_s)}}\label{eta1},\\
    \eta_n^{(2)}&=\frac{\Psi^{(1)T}_{n}(\tau_s)i\sigma_2\left[\gamma^{(1)}_n\Gamma^{(1)}_n(\tau_s)+\gamma^{(2)}_n\Gamma^{(2)}_n(\tau_s)\right]}{\Psi^{(1)T}_{n}(\tau_s)i\sigma_2\Psi^{(2)}_{n}(\tau_s)}.\label{eta2}
\end{align}
This completes the construction of the complete set of positive norm field modes.

Finally, quantization is achieved by promoting each Fourier coefficient $a_n$ in Eq.~\eqref{fieldexp} to an operator subjected to the canonical commutation relation $[a_n,a^\dagger_{n'}]=\delta_{n,n'}$. Furthermore, the vacuum state, $|0\rangle$, is defined by the kernel condition $a_n|0\rangle=0$ for all $n$. Thus,
\begin{equation}
\chi(t,x)=e^{-i\mu t}\sum_n\left[a_n\Phi_{n,1}(t,x)+a_n^\dagger\Phi^*_{n,2}(t,x)\right]\label{quantumexp}
\end{equation}
is the operator-valued distribution that models the quantum fluctuations on top of the background condensate.

\subsection{Instantaneous vacuum states in the interacting period}

The quantum field expansion \eqref{quantumexp}, or, equivalently, \eqref{fieldexp}, also carries information about instantaneous (quasiparticle) vacuum states in the interacting period ($t>\tau_s$), which are obtained if the quantization is performed without the knowledge of the system history. In order to identify an instantaneous quasiparticle vacuum, we need only to find a complete set of positive norm field modes in the interaction period. For $n\neq0$, we find that all $e^{ik_n x}\Psi^{(1)}_n(t)$
have norm $1$ and are eigenfunctions of $i\partial_t$. For $n=0$, though, the field modes are more intricate. We note that
\begin{align}
    \Pi_0&=\left(\begin{array}{cc}
      1  \\
      -1
    \end{array}\right),\\
    \tilde{\Pi}_0&=\frac{1}{2}\left(\begin{array}{cc}
      1  \\
      1
    \end{array}\right)-it\left(\begin{array}{cc}
      1  \\
      -1
    \end{array}\right),
\end{align}
are two independent solutions of Eq.~\eqref{BdGspinor}, such that $\langle\Pi_0,\Pi_0\rangle=\langle\tilde{\Pi}_0,\tilde{\Pi}_0\rangle=0$, and $\langle\Pi_0,\tilde{\Pi}_0\rangle=\ell$. Also, $\tilde{\Pi_0}$ is not an eigenfunction of $i\partial_t$, which is the mathematical expression for the breakdown of the time translation symmetry. Following \cite{Cesar}, the field mode
\begin{equation}
    \Psi^{(1)}_0=\frac{\alpha}{\sqrt{\ell}}\Pi_0+\frac{1}{\sqrt{\ell}}\left(\frac{1}{2\alpha}+i\beta\right)\tilde{\Pi}_0,
\end{equation}
for $\alpha>0$, and $\beta\in\mathbb{R}$, has norm $1$. Each pair $(\alpha,\beta)$ characterizes a distinct instantaneous vacuum state for the theory. Naturally, this choice does not interfere with the physical properties of the system when it evolves from the vacuum defined by the operators $a_n$.
Therefore, for all $n$ and for $t>\tau_s$, we find
\begin{equation}
    \Phi_n(t,x)=e^{ik_n x}[\eta^{(1)}_n\Psi^{(1)}_n(t)+\eta^{(2)}_n\sigma_1\Psi^{(1)*}_n(t)],
\end{equation}
with $\eta^{(1)}_n$, $\eta^{(2)}_n$ given by Eqs.~\eqref{eta1} and \eqref{eta2} for $n\neq0$, and
\begin{align}
    \eta_0^{(1)}&=\alpha+\frac{1+i\tau_s}{2}\left(\frac{1}{2\alpha}-i\beta\right),\\
    \eta_0^{(2)}&=\alpha-\frac{1+i\tau_s}{2}\left(\frac{1}{2\alpha}+i\beta\right),
\end{align}
for $n=0$. Note that the property $1=\langle\Phi_n,\Phi_n\rangle$ implies that
\begin{equation}
    |\eta^{(1)}_n|^2-|\eta^{(2)}_n|^2=1,\label{bogo}
\end{equation}
for all $n$. Finally, it follows from Eq.~\eqref{fieldexp} that
\begin{equation}
    \Phi(t,x)=\sum_n\left[b_ne^{ik_n x}\Psi^{(1)}_n(t)+b^\dagger_ne^{-ik_n x}\sigma_1\Psi^{(1)*}_n(t)\right],\label{quantumqp}
\end{equation}
where we defined the destruction operator
\begin{equation}
    b_n=a_n\eta^{(1)}_n
+a^\dagger_{-n}\eta^{(2)*}_{n},
\end{equation}
which, in view of Eq.~\eqref{bogo}, satisfies $[b_n,b^\dagger_{n'}]=\delta_{n,n'}$. The above equation expresses the fact that the two quantum field expansions are connected via a Bogoliubov transformation, and the quasiparticle vacuum state defined by the $b_n$, i.e., the state $|0\rangle_{\rm qp}$ such that $b_n|0\rangle_{\rm qp}=0$ for all $n$, gives rise to a distinct quantum field theory. In particular, note that $|0\rangle_{\rm qp}$ depends on the parameters $(\alpha,\beta)$. In the next section we use the quantum expansion thus constructed to calculate how the energy in the system is distributed as the condensate evolves.

\section{Consequences of the quantum quench}
\label{section5}

\subsection{Quantum depletion}

We now have the necessary machinery to discuss the energy distribution among the various parts of the condensate. We start from the condensate depletion, as it will be used later on to showcase the system energy.

For the sake of completeness and due to its crucial importance to the backreaction analysis in BECs, we consider first the quantum depletion for a condensate {\it known} to be in the interaction regime and in some quasiparticle vacuum. By defining $\rho_{\chi,{\rm qp}}=\langle\chi^\dagger\chi\rangle_{\rm qp}$ taken with respect to $|0\rangle_{\rm qp}$ , we find, by means of Eq.~\eqref{quantumqp} that
\begin{equation}
    \ell\rho_{\chi,{\rm qp}}=\left|\Psi^{(1)}_{0,2}\right|^2+2\sum_{n=1}^{\infty}\left|\Psi^{(1)}_{n,2}\right|^2,
\end{equation}
where we separated off the contribution from the $n=0$ mode. We note that the sum in the above equation is time-independent in view of Eq.~\eqref{psidef}, whereas the first term is not. Furthermore, the first term depends on the particular choice of the values of $\alpha$ and $\beta$, and some physical criterion must be imposed in order to select a vacuum state. For the particular case of BECs, there is a preferred choice, given by the unique vacuum state such that the number of depleted particles is minimum at $t=t_0>\tau_s$. This idea was proposed in \cite{Cesar}, and an analogous procedure in the context of semiclassical cosmology was explored recently in \cite{Sandro}. A straightforward manipulation leads to
\begin{align}
    \alpha &=  \frac{\sqrt{1 + 4 t_0^2}}{2},\\
    \beta &= -\frac{t_0}{\alpha},
\end{align}
and
\begin{equation}
    \ell\rho_{\chi,{\rm qp}}=(t-t_0)^2+2\sum_{n=1}^{\infty}\left|\Psi^{(1)}_{n,2}\right|^2.
\end{equation}

The important feature of this result is that $\rho_{\chi,{\rm qp}}$ grows with $t^2$, and this behavior occurs for all values of $\alpha$ and $\beta$. Nevertheless, we stress that it is not clear how to experimentally prepare a condensate in one of such vacuum states. We cite \cite{Diehl2008} for a related discussion. 

\begin{figure}[h!]
\center
\includegraphics[width=0.45\textwidth]{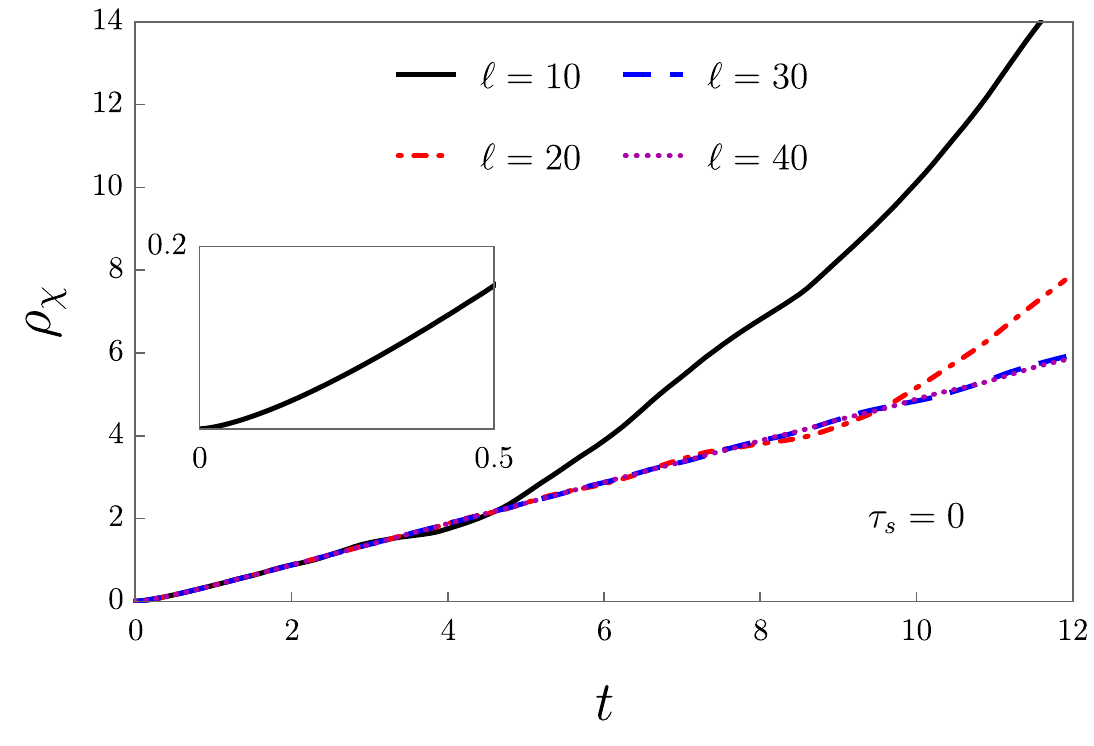}
\caption{Quantum depletion for several condensate sizes. Here the switching time, $\tau_s$, is zero, characterizing a sudden transition from the non-interacting to the interacting regime. Notice the same linear growth law presented by all curves initially, followed by a transition to a different power law occurring as function of the system size. Inset: Quantum depletion for $\ell=10$, revealing an initial $t^2$ behavior. We recall that all time scales are presented in units of $1/\Delta\mu$, and all distance scales, in units of $\xi_0$.
}
\label{fig1}
\end{figure}

A more realistic description is obtained from the vacuum $|0\rangle$, defined by the $a_n$ of the expansion \eqref{quantumexp}. The quantum depletion in this case reads
\begin{equation}
    \rho_{\chi}=\left|\Phi^{(1)}_{0,2}\right|^2+2\sum_{n=1}^{\infty}\left|\Phi^{(1)}_{n,2}\right|^2,
\end{equation}
where now all terms in the above depend on time. Similarly to what occurs to $\rho_{\chi,{\rm qp}}$, we find, in view of Eq.~\eqref{zeromode}, that in the interaction period, $t>\tau_s$, the mode with $n=0$ is excited by the continuous switching in such a way that
\begin{equation}
    \ell\left|\Phi^{(1)}_{0,2}\right|^2=\left(t-\frac{\tau_s}{2}\right)^2,
\end{equation}
reproducing the $t^2$ growth as $t\rightarrow\infty$. However, as depicted in Fig.~\ref{fig1}, modes with $n\neq0$ are excited in such a way that the depletion growth is not purely quadratic.
Inspection of Fig.~\eqref{fig1} reveals that for the sudden switching case, $\tau_s=0$, depletion grows linearly with time regardless of the system size. This behavior was observed also in the bulk of a finite size 1D condensate in \cite{RibeiroPRA}. The novelty here is the transition from the linear behavior after some time depending on the system size. This transition shows that the dynamics of the condensate depletion can show intricate behavior even for spatially homogeneous configurations.

\begin{figure}[h!]
\center
\includegraphics[width=0.45\textwidth]{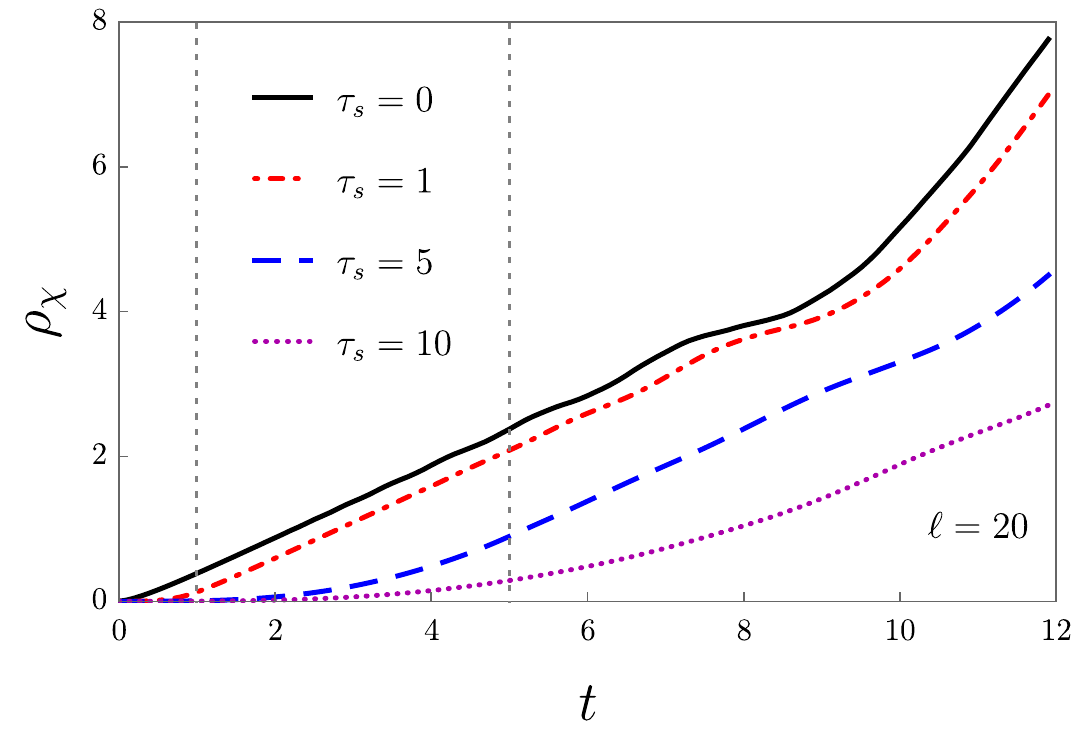}
\caption{Quantum depletion for $\ell=20$ and several switching times $\tau_s$. The dashed and dot-dashed curves show that the quantum depletion undergoes linear growth after $t=\tau_s$, thus suggesting that this behavior might be present in the bulk all 1D condensates undergoing ``stationary'' ($g$ constant) phase diffusion.  
}
\label{fig2}
\end{figure}

Figure \ref{fig2} presents several depletion profiles as function of the switching time $\tau_s$ for condensates with $\ell=20$, which can be used to investigate the influence of the sudden quantum quench on the linear growth behavior observed in Fig.~\ref{fig1}.
From the profiles presented in the figure it is possible to verify that the linear depletion growth is not caused by the sudden transition to the interacting regime. Indeed, the dot-dashed curve, which corresponds to a switching time of $\tau_s=1$, clearly shows that the quantum depletion reaches the linear growth regime after $t=\tau_s$.

\subsection{Condensate energy}

We now consider the energy balance in the condensate after the interactions are turned on. Even though the backreaction analysis is simple for this condensate profile, with $\rho_\zeta=-\rho_\chi$ and $J_\chi=J_\zeta=0$ being the solutions to the backreaction equations, the energy density stored in the depleted particle cloud, $\mathcal{E}_\chi$, is not equal to the energy given to the background condensate, $\mathcal{E}_\zeta$, and their behavior are fairly rich and distinct.

We recall that due to the absence of particle fluxes in this condensate configuration and the fact that $\rho_\chi=-\rho_\zeta$, the conservation of energy in this system assumes the form
\begin{align}
    \partial_t(\mathcal{E}_\chi+\mathcal{E}_\zeta)=\frac{G^{(2)}}{2}(\partial_tg),\label{energyconaux}
\end{align}
and thus, in view of Eq.~\eqref{eqforg}, work is done on the system during the switching period, when energy is not conserved. 

\begin{figure}[h!]
\center
\includegraphics[width=0.45\textwidth]{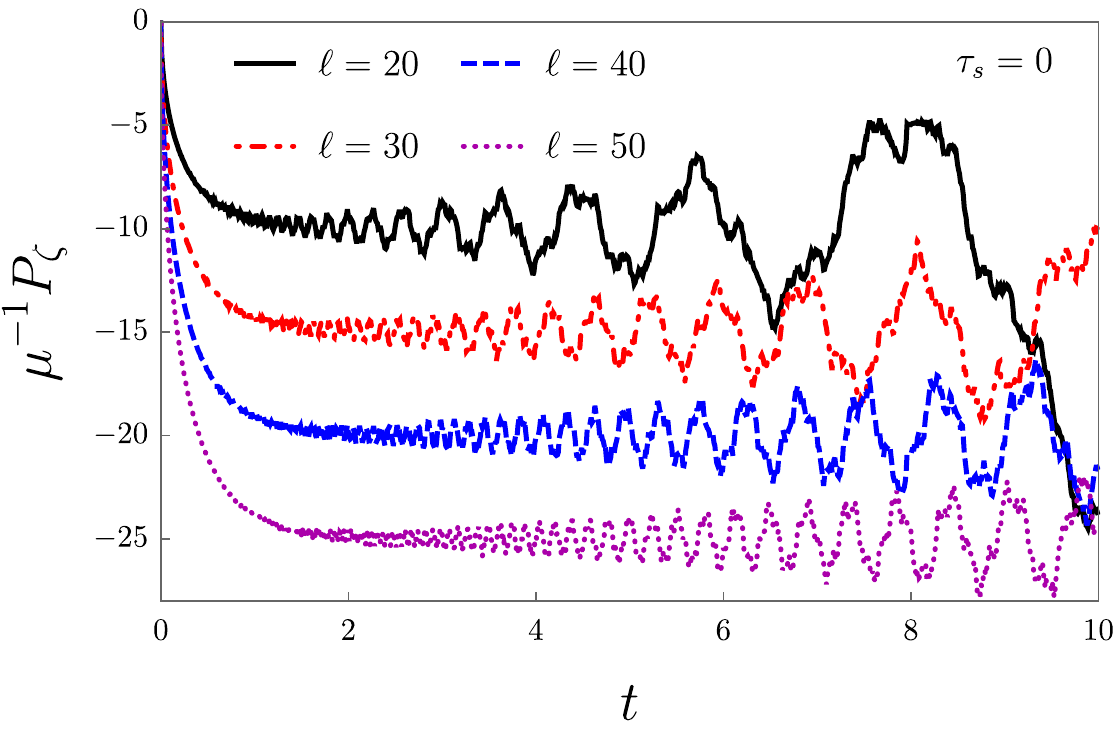}
\caption{Power transferred to the background condensate by the quantum fluctuations for several condensate sizes, and in the sudden transition regime, $\tau_s=0$. The curves reveal another curious aspect of condensate evolution, which is the non-monotonic transfer of energy to the condensate.  
}
\label{fig3}
\end{figure}

Let us analyze first the energy density given to the condensate background, $\mathcal{E}_\zeta$. It follows from the definition of the chemical potential \eqref{chemical} and Eq.~\eqref{energyzeta} that
\begin{equation}
    \mathcal{E}_\zeta=-\mu\rho_\chi,
\end{equation}
which has a simple interpretation: As $\mu$ is by definition the system energy per particle, $-\mathcal{E}_\zeta>0$ is the energy removed from the system per particle per unit healing length. 
Also, because all terms in the above do not depend on $x$, it is instructive to discuss the power {\it transferred to} the background condensate, $P_\zeta=\int\d x\partial_t\mathcal{E}_{\zeta}$. We find that
\begin{align}
    P_{\zeta}=-\mu\ell\partial_t\rho_\chi,
\end{align}
and Fig.~\ref{fig3} depicts 
$P_{\zeta}$ for several condensate sizes in the sudden transition configuration. We recall that the physical interpretation of $P_{\zeta}$ is straightforward: It is the energy per unit healing length transferred to the background condensate by the quantum fluctuations and development of the depletion cloud. Figure \ref{fig3} shows that for all considered cases, as the particles start to interact, energy extraction from the condensate initially occurs in an accelerated manner and then reaches a oscillatory behavior around a constant value. 

We also note that the amplitude of the oscillations grows with time and we interpret that as an indicative of a dynamical instability in the condensate evolution, which is expected as the condensate phase degrades with time.  
\begin{figure}[h!]
\center
\includegraphics[width=0.45\textwidth]{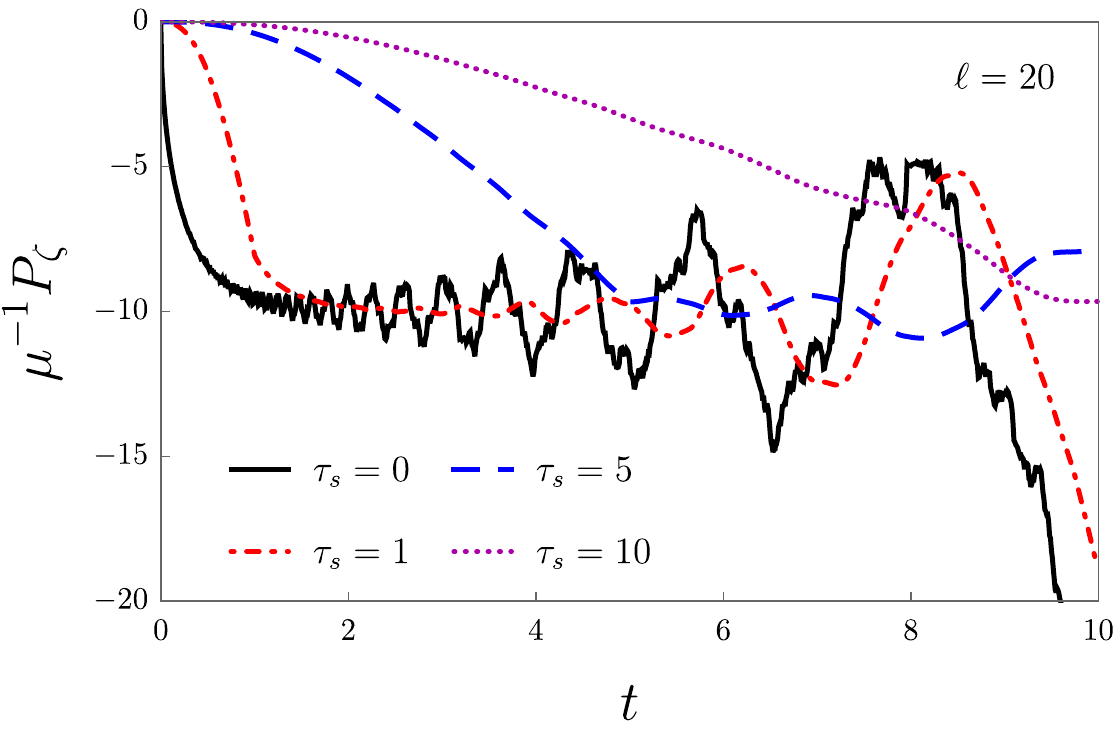}
\caption{Power transferred to the background condensate by the quantum fluctuations as function of the switching time $\tau_s$ and for a fixed condensate size $\ell=20$. It is possible to conclude that the effect of the continuous switching mechanism is to diminish the initial acceleration of the energy extraction rate. 
}
\label{fig4}
\end{figure}
Figure \ref{fig4} illustrates the dependence of the power $P_\zeta$ as function of the switching time $\tau_s$, and it is possible to conclude that the overall effect of the continuous switching mechanism is to increase the amount of time taken for the energy extraction rate to reach the oscillatory regime revealed by the plots in Fig.~\ref{fig3}.

We now consider the final ingredient of the system energy, $\mathcal{E}_{\chi}$. From the definition \eqref{energychi} we find that $\mathcal{E}_{\chi}$ can be written as
\begin{equation}
    \mathcal{E}_{\chi}=-\mathcal{E}_{\zeta}+(\Delta\mu)\left[\frac{1}{2}\langle(\partial_x\chi^\dagger)\partial_x\chi\rangle+\frac{g}{2g_0\rho_0}G^{(2)}\right].\label{energyaux}
\end{equation}
The first term in Eq.~\eqref{energyaux} has a clear interpretation: It is the energy per particle per unit healing length stored in the depletion cloud, and, consequently, it is completely determined by the system depletion, whose physical properties were already discussed in the above. However, differently from $\mathcal{E}_\zeta$, $\mathcal{E}_\chi$ contains an extra part, which is the energy stored on the depleted cloud due to the work done by the external agents, as shown by Eq.~\eqref{energyconaux}. 

\begin{figure}[h!]
\center
\includegraphics[width=0.45\textwidth]{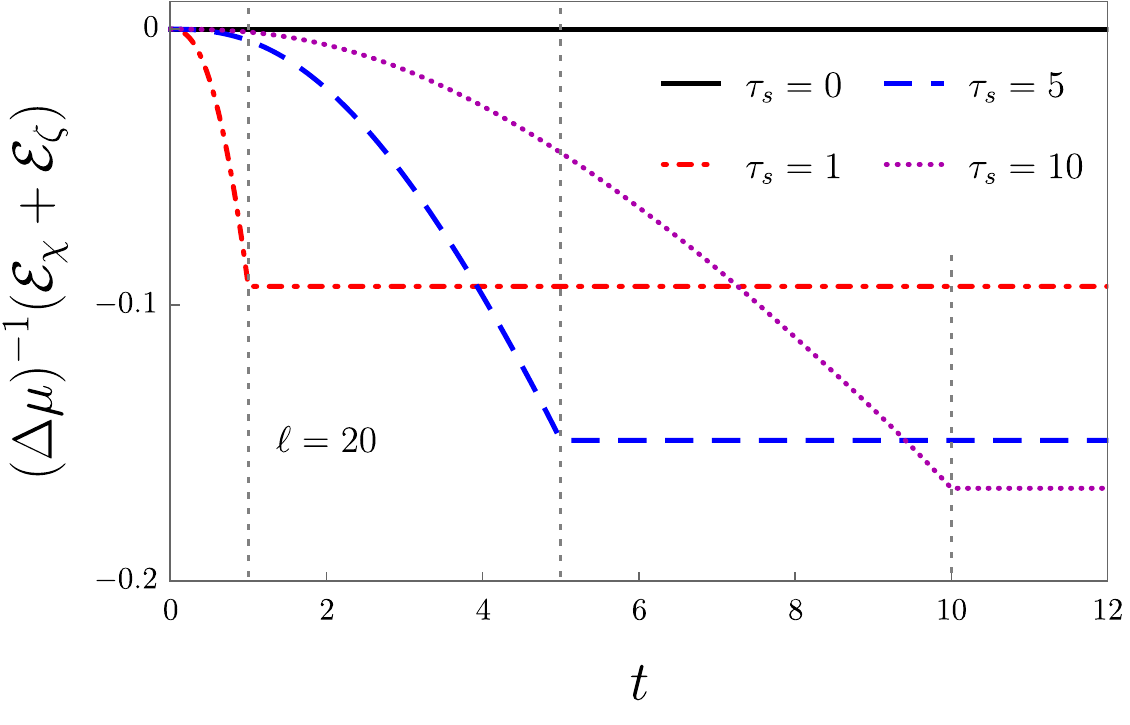}
\caption{Energy per unit healing length acquired by the depleted cloud due to the work done by external forces, for several values of $\tau_s$. As ensured by energy conservation, $\mathcal{E}_{\chi}+\mathcal{E}_{\zeta}$ can only depend on time if $\tau_s>0$, in which case $\partial_tg\neq0$. 
}
\label{fig5}
\end{figure}

In Fig.~\ref{fig5} we depict $\mathcal{E}_{\chi}+\mathcal{E}_\zeta$ for several values of the switching time for a condensate with $\ell=20$.
Equation \eqref{energyconaux} shows that $\mathcal{E}_{\chi}+\mathcal{E}_{\zeta}$ can only vary with time if $\partial_tg\neq 0$, i.e., during the switching period, what is confirmed in Fig.~\ref{fig5}. Furthermore, because $\mathcal{E}_{\chi}+\mathcal{E}_{\zeta}<0$ for the condensate configurations of Fig.~\ref{fig5}, we also conclude that the effect of the external forces is to diminish the energy of the depleted particles.

We finish this analysis with a comment about the possibility of probing the effects discussed in the above. It is possible to measure depletion, as was done, for instance, in \cite{lopes}. This can be used to determine $\mathcal{E}_{\zeta}$. Moreover, Equation \eqref{energyconaux} teaches us that one can calculate $\mathcal{E}_{\chi}+\mathcal{E}_{\zeta}$ from measurements of $G^{(2)}$. The experimental determination of $G^{(2)}$ was performed, for example, in \cite{Jeff2019} as a means to detect the analogue Hawking radiation. Therefore, it is possible, with currently available experimental methods, to determine both $\mathcal{E}_{\chi}$ and $\mathcal{E}_{\zeta}$.

\section{Final Remarks}
\label{section6}

In this work we considered the conservation of energy and momentum in the context of quantum backreaction in Bose-Einstein condensates. By employing the Bogoliubov expansion and the canonical stress tensor, $T^{\mu}_{\ \nu}$, as a measure of the energy and momentum in the system, the contributions to $T^{\mu}_{\ \nu}$ coming from quantum fluctuations and condensate corrections were identified as well as the associated conservation laws. 

The backreaction equations were exactly solved for a family of homogeneous condensates in a ring geometry during a transition between a non-interacting phase and an interacting regime. The solution to the backreaction problem was then used to study the energy balance in the system, revealing nuances of the condensate dynamics that occur even for simplified condensate configurations.

We finish this work with some remarks about the considered condensate model. All particle fluxes vanish for the adopted homogeneous background condensate, rendering the law of conservation of momentum trivially satisfied. As anticipated in the introduction, one problem of great interest is the study of the backreaction equations for black hole analogues, for which both conservation laws, for energy and momentum, are expected to showcase a much richer hall of physical phenomena. However, due to the complexity of the backreaction equations for these cases, exact solutions like the one built here are in general not accessible, and the solutions should be built numerically. 

Also crucial for the backreaction analysis is the control of the condensate true state at a given instant. We have shown that all the results just found for the non-stationary condensate configuration are completely different from results that would be obtained if some instantaneous vacuum state were considered. This might be a source of discrepancy between experiments and simulations with these systems, and could offer, for instance, some light over the discrepancy observed in \cite{Ross2022}.  

Finally, we call attention to the use of the canonical stress tensor as a measure of the system energy and momentum. Although the conservation laws and the definitions of the system energy and momentum are correct, the definition of the system true energy flux and stress might be different. We recall that the fundamental stress tensor for a given field theory is the one that enters Einstein's equations \cite{wald2022}, and therefore it must be symmetric. An interesting development of the results obtained here would be to consider what differences are obtained for a relativistic condensate in curved spaces.

\section*{Acknowledgements} 
C.C.H.R. would like to thank the Funda\c{c}\~ao de Apoio \`a Pesquisa do Distrito
Federal (grant 00193-00002051/2023-14) for supporting this work. 

\appendix
\bibliography{qgav3.bib}
\end{document}